\def\dmsol{\Delta m^2_\odot}
\def\dmatm{\Delta m^2_@}

\documentclass{elsart}
\usepackage{amssymb}
\usepackage{amsmath}
\usepackage{graphicx}
\journal{Physics Letters B}
\begin{document}

\begin{frontmatter}
\title{Removing Ambiguities in the Neutrino Mass Matrix}

\author[Lisbon]{G. C. Branco},
%\thanks[branco]{gbranco@alfa.ist.utl.pt}
\author[Lisbon]{R. Gonz{\'a}lez Felipe},
%\thanks[gonzalez]{gonzalez@gtae3.ist.utl.pt}
\author[Lisbon]{F. R. Joaquim},
%\thanks[filipe]{filipe@gtae3.ist.utl.pt}
\author[Tokyo]{T. Yanagida}
%\thanks[yanagida]{tsutomu.yanagida@cern.ch}

\address[Lisbon]{Departamento de F{\'\i}sica and Grupo de F{\'\i}sica de Part{\'\i}culas (GFP),
Instituto Superior T{\'e}cnico, Av.Rovisco Pais, 1049-001 Lisboa, Portugal}
\address[Tokyo]{Department of Physics, University of Tokyo, Tokyo 113-0033, Japan}

\begin{abstract}
We suggest that the weak-basis independent condition $\det\,(M_\nu)=0$ for the
effective neutrino mass matrix can be used in order to remove the ambiguities
in the reconstruction of the neutrino mass matrix from input data available
from present and future feasible experiments. In this framework, we study the
full reconstruction of $M_\nu$ with special emphasis on the correlation between
the Majorana CP-violating phase and the various mixing angles. The impact of
the recent KamLAND results on the effective neutrino mass parameter is also
briefly discussed.

\end{abstract}
%\begin{keyword}
%Neutrino masses and mixing, CP violation
%\end{keyword}
\end{frontmatter}

\section{Introduction}

Neutrino masses and mixings are most likely described by a Majorana mass matrix
which naturally arises in the context of the standard electroweak gauge theory,
with the implicit assumption that $B-L$ is violated at a high-energy scale. The
smallness of neutrino masses is then elegantly explained by the seesaw
mechanism \cite{Yanagida:1979}. The $3 \times 3$ complex symmetric Majorana
neutrino mass matrix $M_\nu$ contains nine physical parameters, while realistic
experiments can determine only seven independent quantities. This leads to the
wretched situation that no set of feasible experiments can fully determine the
neutrino mass matrix. This basic observation has encouraged Frampton, Glashow
and Marfatia \cite{Frampton:2002yf} to propose that the neutrino mass matrix
$M_\nu$ contains texture zeros, in order to reduce the number of free
parameters. However, the presence of zeros in $M_\nu$ crucially depends on the
weak basis one chooses. Therefore, it is desirable to consider
basis-independent constraints on the neutrino mass matrix.

In this letter we address the question of whether it is possible to achieve an
appropriate reduction of parameters through the introduction of a weak-basis
independent condition. We propose the basis independent condition that the
determinant of the neutrino mass matrix vanishes, that is $\det\,(M_\nu)=0$.
Since this condition gives two constraints on the parameters, the neutrino mass
matrix has now just 7 parameters which can be fully determined  by future
feasible experiments. We note that the fact that $\det\,(M_\nu)=0$ implies the
elimination of two parameters has to do with the assumed Majorana nature of
neutrinos. In contrast, if one imposes the condition $\det\,(M_u)=0$ in the
quark sector, one looses only one parameter in the electroweak sector, since
the CKM matrix continues having four parameters even in the limit $m_u=0$. On
the other hand, the condition $m_u=0$ allows to remove another parameter from
the theory, since it implies $\bar{\theta}=0$ ($\bar{\theta}$ is the
coefficient of $F_{\mu\nu} \tilde{F}_{\mu\nu}$), thus providing a possible
solution to the strong CP problem \cite{Peccei:1988ci}. It is also interesting
that the Affleck-Dine scenario for leptogenesis \cite{Affleck:1984fy} requires
the mass of the lightest neutrino to be $m_1 \simeq 10^{-10}$~eV
\cite{Asaka:2000nb,Fujii:2001sn}, which leads practically to our condition
$\det\,(M_\nu)=0$. Furthermore, such an extremely small mass may be explained
by a discrete $Z_6$ family symmetry \cite{Fujii:2001sn}. Within the framework
of the seesaw mechanism, the study of leptonic CP violation at high energies
and its relation to the neutrino mass spectrum could have profound cosmological
implications, for instance in the generation of the observed baryon asymmetry
of the universe through leptogenesis \cite{Fukugita:1986hr,Branco:2001pq}.
However, in this letter we will restrict ourselves to low energies and
therefore our analysis remains valid independently of the high energy origin of
the effective neutrino mass matrix.

In the framework where $\det\,(M_\nu)=0$, we study the full reconstruction of
$M_\nu$ with special emphasis on the correlation between the Majorana
CP-violating phase and the various mixing angles. We also discuss how future
neutrinoless double beta decay experiments could invalidate the assumed
weak-basis independent condition.

\section{Reconstruction of the neutrino mass matrix}

Let us start by summarizing the present data on neutrino mass-squared
differences and mixing angles, obtained from the evidence for neutrino
oscillations in atmospheric, solar and reactor neutrino experiments. Assuming
two-neutrino mixing and dominant $\nu_{\mu} \rightarrow \nu_{\tau}\,
(\bar{\nu}_{\mu} \rightarrow \bar{\nu}_{\tau})\,$ oscillations, the atmospheric
neutrino data obtained from Super-Kamiokande experiments yields at 99.73 \%
C.L. \cite{Sobel:tj}:
\begin{align} \label{atmdata}%
1.5 \times 10^{-3}\; \text{eV\,}^2 \leq \dmatm & \leq 5.0 \times 10^{-3}\;
\text{eV\,}^2\, , \nonumber\\
\sin^2 2 \theta_@ & > 0.85\, ,
\end{align}
with the best-fit values $(\dmatm)_{BF}=2.5 \times 10^{-3}\; \text{eV\,}^2\,$
and $(\sin^2 2 \theta_@)_{BF}=1\ $. On the other hand, global neutrino analyses
of the solar neutrino data \cite{Ahmad:2001an} under the assumption of $\nu_{e}
\rightarrow \nu_{\mu,\tau}\,$ oscillations/transitions favors the LMA MSW solar
solution with
\begin{align} \label{soldata}%
2.2 \times 10^{-5}\; \text{eV\,}^2 \leq \dmsol & \leq 2.0 \times 10^{-4}\;
\text{eV\,}^2\, , \nonumber\\
0.18 \lesssim \sin^2 \theta_\odot & \lesssim 0.37\, ,
\end{align}
and the best-fit values $(\dmsol)_{BF}=5 \times 10^{-5}\; \text{eV\,}^2\,$ and
$(\sin^2 \theta_\odot)_{BF}=0.25\ $. Finally, the reactor neutrino data
obtained from the CHOOZ experiment \cite{Apollonio:1999ae} puts an upper bound
on the leptonic mixing matrix element $U_{e3}$. The combined three-neutrino
oscillation analyses of the solar, atmospheric and reactor data imply at 99.73
\% C.L. \cite{Gonzalez-Garcia:2000sq,Fogli:2002pb}:
\begin{align} \label{reacdata}%
 |U_{e3}| < 0.22\, ,
\end{align}
with a best-fit value of $(|U_{e3}|)_{BF} \simeq 0.07$ found in
\cite{Gonzalez-Garcia:2000sq}.

Another important input information comes from neutrinoless double $\beta$
decay experiments, which could provide us with evidence for non-vanishing
CP-violating Majorana phases, and thus, for the Majorana nature of massive
neutrinos \cite{Bilenky:2001rz}. For Majorana neutrinos with masses not
exceeding a few MeV, the amplitude of this process is proportional to the
so-called effective Majorana mass parameter $m_{ee} = |(M_{\nu})_{11}|\ $.
Although no evidence for $(\beta\beta)_{0\nu}$ decay has been found so far,
rather stringent upper bounds have been obtained. In particular, the $^{76}$Ge
Heidelberg-Moscow experiment has reported the limit $m_{ee} < 0.35$~eV at 90 \%
C.L. and the IGEX collaboration, $m_{ee} < (0.33-1.35)$~eV at 90 \% C.L. . A
considerable higher sensitivity is expected in future experiments. For
instance, values of $m_{ee} \simeq 5.2 \times 10^{-2}$ (EXO), $m_{ee} \simeq
3.6 \times 10^{-2}$ (MOON) and $m_{ee} \simeq 2.7 \times 10^{-2}$ (CUORE) are
planned to be achieved \cite{Staudt:qi}.

As far as the CP-violating Dirac phase is concerned, there is at present no
experimental information on its value. However, neutrino factories will in
principle be able to measure $\text{Im}\,(U_{e2} U_{\mu3} U_{\mu2}^\ast
U_{e3}^\ast)$ and thus determine the Dirac phase in any specific
parametrization of the leptonic mixing matrix $U$.

In our framework where $\det\,(M_\nu)=0$, the neutrino mass matrix is
characterized by seven parameters and therefore one can use the above described
seven inputs from experiment to fully reconstruct $M_\nu$ in the weak basis
where the charged lepton mass matrix is diagonal and real. In this basis,
$M_\nu$ can be written as:
\begin{align} \label{Mnuflavor}
M_{\nu}=U^\ast\, \text{diag}\, (m_1 , m_2\, e^{i\alpha_2} , m_3\,
e^{i\alpha_3})\, U^\dagger\,,
\end{align}
where $m_i$ are the moduli of the light neutrino masses, $\alpha_i$ are the two
Majorana phases \cite{Bilenky:1980cx} and $U$ is the MNS neutrino mixing
matrix, which we choose to parametrize in the form:
\begin{align}
U  = \left(
\begin{array}{ccc}
c_{12} c_{13} & s_{12} c_{13} & s_{13}  \\
-s_{12} c_{23} - c_{12} s_{23} s_{13} e^{ i \delta} & -c_{12} c_{23} - s_{12}
s_{23} s_{13} e^{ i \delta} &
s_{23} c_{13}e^{i \delta} \\
s_{12} s_{23} - c_{12} c_{23} s_{13} e^{ i \delta} & -c_{12} s_{23} - s_{12}
c_{23} s_{13} e^{ i \delta} & c_{23} c_{13}e^{i \delta}
\end{array}
\right)\,, \label{Unu}
\end{align}
where $c_{ij} \equiv \cos \theta_{ij}\ $, $s_{ij} \equiv \sin \theta_{ij}\ $,
$0 \leq \theta_{ij} \leq \pi/2$ and $\delta$ is the CP-violating Dirac phase.
The above parametrization turns out to be more convenient in the analysis of
the effective mass parameter $m_{ee}\,$, since in this case this matrix element
depends only on the Majorana phases $\alpha_i$ and not on the Dirac phase
$\delta$. If one uses instead the standard parametrization \cite{Hagiwara:pw},
$U_\nu = U\, \text{diag}\,(1,1,e^{-i\delta})$, then the phase $\delta$ would
enter in the combination $\alpha_3-2\delta$.

The condition $\det\,(M_\nu) = 0$ together with the experimental constraints on
$\dmsol$ and $\dmatm$ imply that only one neutrino can have a vanishing mass.
By identifying the indexes 12 and 23 with the solar and atmospheric neutrinos,
respectively, two possible scenarios can be distinguished. In the first case,
the diagonal matrix in Eq.~(\ref{Mnuflavor}) is of the form
\begin{align}
\text{diag}\, (0 , m_2\, e^{i\alpha} , m_3)\quad \text{(Case I)},
\end{align}
while in the second one, the above matrix is
\begin{align}
\text{diag}\, (m_1 , m_2\, e^{i\alpha} ,0) \quad \text{(Case II)},
\end{align}
with the relevant Majorana phase given in both cases by $\alpha = \alpha_2 -
\alpha_3$.

Since it is for the matrix element $m_{ee}$ that we have direct experimental
access, from now on we will restrict our analysis to the implications of our
assumptions in the determination of this parameter and, consequently, of the
Majorana phase $\alpha$ \cite{Frigerio:2002rd}.

\subsection{Case I: Standard hierarchy}

In this case $m_2 =\sqrt{\dmsol}$ and $m_3=\sqrt{\dmatm+\dmsol}$ and it follows
from Eq.~(\ref{Mnuflavor}) that
\begin{align}
m_{ee}^2 \equiv |(M_\nu)_{11}|^2= & \, m_2^2\,|U_{12}|^4+m_3^2\,|U_{13}|^4
\nonumber\\
+ & 2\,m_2\,m_3\,|U_{12}|^2\,|U_{13}|^2\,\cos \alpha\,.
\end{align}
In the parametrization (\ref{Unu}) we have then
\begin{align} \label{meecaseI}
m_{ee}^2 =
m_2^2\,c_{13}^4\,s_{12}^4+m_3^2\,s_{13}^4+2\,m_2\,m_3\,c_{13}^2\,s_{13}^2\,s_{12}^2
\cos \alpha\,,
\end{align}
with the upper and lower bounds given by
\begin{align}
& m_{ee}^{\rm upper} = |m_2\,s_{12}^2+s_{13}^2\,(m_3-m_2\,s_{12}^2)|\,, \label{meeupper}\\
& m_{ee}^{\rm lower} = |m_2\,s_{12}^2-s_{13}^2\,(m_3+m_2\,s_{12}^2)|\,,
\label{meelower}
\end{align}
which correspond to $\alpha = 0$ (same CP parity) and $\alpha=\pi$ (opposite CP
parities), respectively. We notice from Eq.~(\ref{meelower}) that cancellations
in $m_{ee}$ can occur if
\begin{align}
\tan^2\,\theta_{13} = \frac{m_2\,s_{12}^2}{m_3} \simeq
\frac{\dmsol\,s_{\odot}^2}{\dmatm}\,.
\end{align}
Using the experimental ranges of Eqs.~(\ref{atmdata})-(\ref{reacdata}), we can
get the limits on $s_{13}$ for such cancellations to occur. We find $0.12
\lesssim U_{e3} \lesssim 0.22$.

\begin{figure}
$$\includegraphics[width=9.5cm]{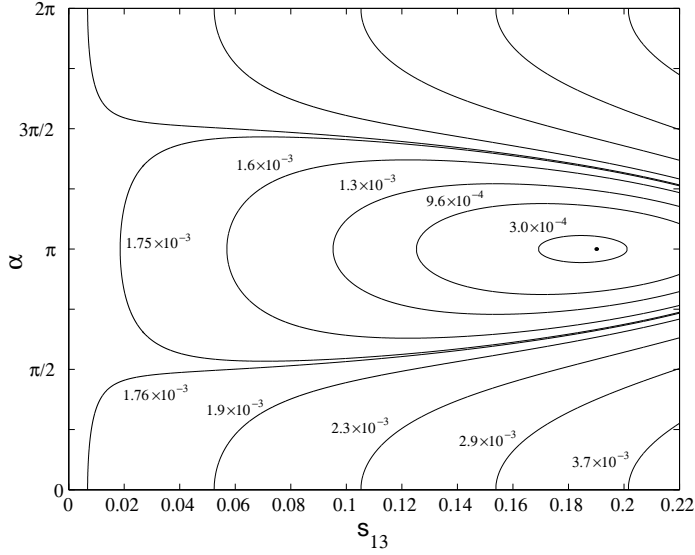}$$ \caption{Contour plots of the Majorana
phase $\alpha$ as a function of $s_{13}$ for different values of the
neutrinoless double $\beta$-decay parameter $m_{ee}\ $. The black dot
corresponds to $m_{ee}=0\ $. The best-fit values of $s_{12},\, \dmsol$ and
$\dmatm$ have been used.} \label{fig1}
\end{figure}

The Majorana phase $\alpha$ can be extracted from Eq.~(\ref{meecaseI}), leading
to:
\begin{align} \label{coscaseI}
\cos \alpha = \frac{m_{ee}^2-m_3^2\,s_{13}^4-m_2^2\,c_{13}^4\,s_{12}^4}
{2\,m_2\,m_3\,c_{13}^2\,s_{13}^2\,s_{12}^2}\,.
\end{align}
In Fig.~\ref{fig1} we present the contour plots of $\alpha$ as a function of
$s_{13}$ for different values of the neutrinoless double $\beta$-decay
parameter $m_{ee}\ $. We consider the best-fit values of $\dmatm$ and
$s_{12},\, \dmsol$ for the LMA MSW solar solution (see Eqs.~(\ref{atmdata}) and
(\ref{soldata})).

\subsection{Case II: Inverted hierarchy}

In this case $m_1 =\sqrt{\dmatm}$ and $m_2=\sqrt{\dmatm+\dmsol}$. For the
effective mass parameter $m_{ee}$ we obtain from Eq.~(\ref{Mnuflavor})
\begin{align}
m_{ee}^2 \equiv |(M_\nu)_{11}|^2= & \, m_1^2\,|U_{11}|^4+m_2^2\,|U_{12}|^4
\nonumber\\
+ & 2\,m_1\,m_2\,|U_{11}|^2\,|U_{12}|^2\,\cos \alpha\,,
\end{align}
which in the parametrization (\ref{Unu}) is equivalent to
\begin{align} \label{meecaseII}
m_{ee}^2 = c_{13}^4\, ( m_1^2\,c_{12}^4+
m_2^2\,s_{12}^4+2\,m_1\,m_2\,s_{12}^2\,c_{12}^2 \cos \alpha )\,.
\end{align}
The upper and lower bounds of $m_{ee}$ are given by
\begin{align}
& m_{ee}^{\rm upper} = c_{13}^2\, (m_1\,c_{12}^2+m_2\,s_{12}^2)\,,\quad
\text{for}\quad \alpha = 0\,,\label{meeupper2}\\
& m_{ee}^{\rm lower} = c_{13}^2\, | m_1\,c_{12}^2-m_2\,s_{12}^2 |\,, \quad \,\,
\text{for}\quad \alpha = \pi \label{meelower2}.
\end{align}
The vanishing of $m_{ee}$ requires $ \tan^2\,\theta_{12} = m_1/m_2 \simeq 1\,.$
Since values of $\tan^2\,\theta_{12} \simeq 1$ are excluded by the present LMA
solar data, such cancellations cannot occur in this case.

\begin{figure}
$$\includegraphics[width=9.5cm]{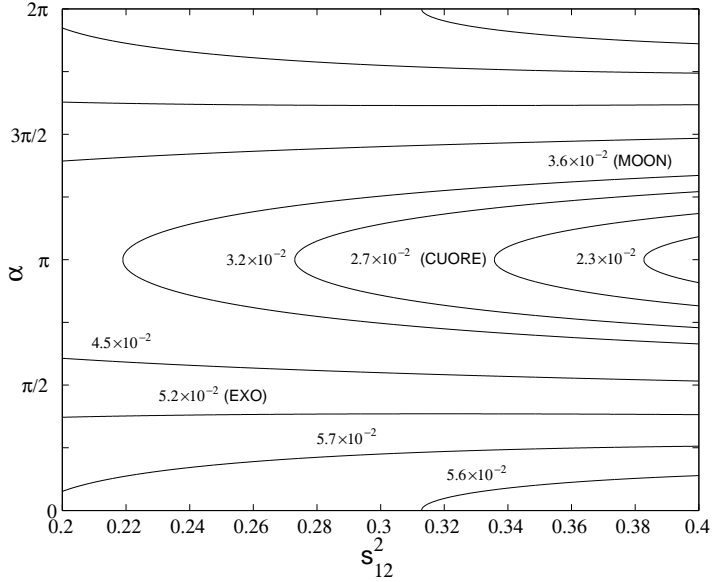}$$ \caption{Contour plots of the Majorana
phase $\alpha$ as a function of $s_{12}^2$ for different values of the
neutrinoless double $\beta$-decay parameter $m_{ee}\ $. The contours
corresponding to the sensitivity of the future MOON, CUORE and EXO
$(\beta\beta)_{0\nu}$-decay experiments have also been included. The best-fit
values of $s_{13},\, \dmsol$ and $\dmatm$ have been used.} \label{fig2}
\end{figure}

From Eq.~(\ref{meecaseII}) we find:
\begin{align} \label{coscaseII}
\cos \alpha =
\frac{m_{ee}^2-m_1^2\,c_{12}^4\,c_{13}^4-m_2^2\,s_{12}^4\,c_{13}^4}
{2\,m_1\,m_2\,s_{12}^2\,c_{12}^2\,c_{13}^4}\,.
\end{align}
We notice that the Majorana phase $\alpha$ is not very sensitive to the small
values of $U_{e3}=s_{13}$ allowed by the present data. On the other hand, the
value of $\alpha$ is more sensitive to the solar mixing angle $\theta_{12}$. In
Fig.~\ref{fig2} we present the contour plots of the Majorana phase $\alpha$ as
a function of $s_{12}^2$ for different values of the neutrinoless double
$\beta$-decay parameter $m_{ee}\ $. The values of $m_{ee}$ are in this case at
the reach of the future $(\beta\beta)_{0\nu}$-decay experiments. For
comparison, the contours corresponding to the sensitivity of the future MOON,
CUORE and EXO experiments are also plotted. We have used the best-fit values of
$s_{13},\, \dmsol$ and $\dmatm$.

\begin{figure}
$$\includegraphics[width=9.5cm]{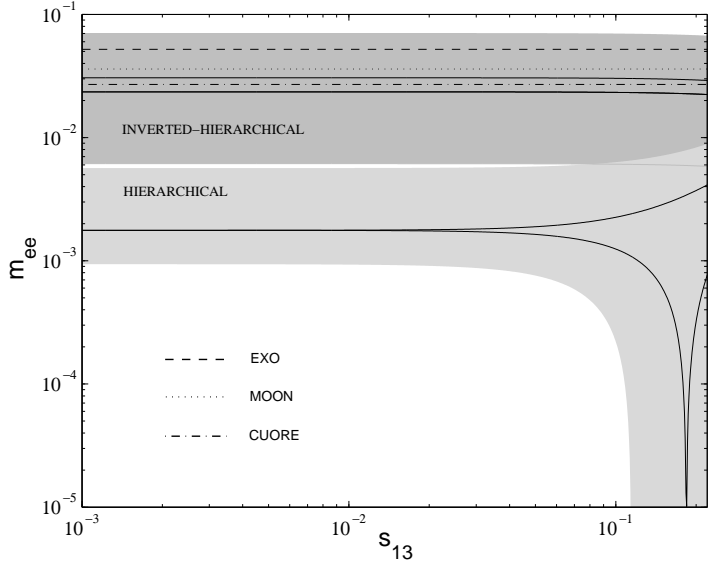}$$ \caption{Allowed regions for the
effective mass parameter $m_{ee}$ in the cases of hierarchical (light grey) and
inverted-hierarchical (dark grey) neutrino mass spectra, assuming
$\det\,(M_\nu)=0$ and taking into account all the available experimental data
for solar, atmospheric and reactor neutrinos as given in
(\ref{atmdata})-(\ref{reacdata}). The areas delimited by the solid lines
correspond to the allowed regions considering the best-fit values of $\dmsol$,
$\dmatm$ and $\theta_\odot$.} \label{fig3}
\end{figure}

In Fig.~\ref{fig3} we present the allowed regions for the effective mass
parameter $m_{ee}$ in the cases of hierarchical (Case I) and
inverted-hierarchical (Case II) neutrino mass spectra, assuming the weak-basis
independent condition $\det\,(M_\nu)=0$. We use the experimental ranges for
solar, atmospheric and reactor neutrinos as given in
Eqs.~(\ref{atmdata})-(\ref{reacdata}). The areas delimited by the solid lines
correspond to the allowed regions if one uses the best-fit values of $\dmsol$,
$\dmatm$ and $\theta_\odot$. This figure also illustrates how future
$(\beta\beta)_{0\nu}$-decay experiments could in principle distinguish the two
cases considered, or even exclude one or both of them. In particular, it is
seen from the figure that if $m_{ee} \lesssim 10^{-3}$ and $U_{e3} \lesssim
0.05$ then the vanishing of the determinant of $M_{\nu}$ is not a viable
assumption. It is also clear that a better knowledge on the mixing angles and
$\Delta m^2$'s is crucial to distinguish between the hierarchical and
inverted-hierarchical cases.

\begin{figure}
$$\includegraphics[width=9.5cm]{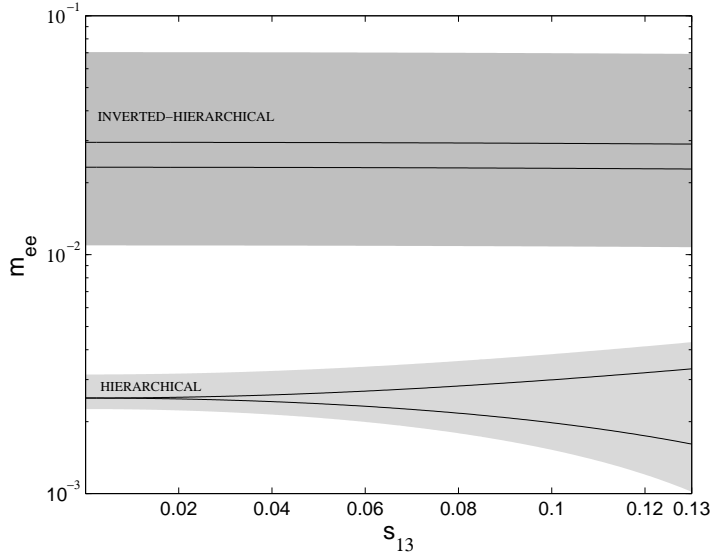}$$ \caption{Allowed regions for the
effective mass parameter $m_{ee}$ in the cases of hierarchical (light grey) and
inverted-hierarchical (dark grey) neutrino mass spectra, assuming
$\det\,(M_\nu)=0$ and taking into account the $\pm 1\sigma$ estimates for
$\Delta m_{12}^2$, $s_{12}^2$ and $s_{13}$ obtained from the global analysis of
the pre-KamLAND and new KamLAND data given in \cite{Fogli:2002au}. The areas
delimited by the solid lines correspond to the allowed regions considering the
best-fit values $\Delta m_{12}^2=7 \times 10^{-5}\,\rm{eV}^2$ and
$s^2_{12}=0.3$.} \label{fig4}
\end{figure}

To conclude our analysis let us briefly comment on the recent results reported
by the Kamioka Liquid Scintillator Antineutrino Detector (KamLAND)
\cite{Eguchi:2002dm}. These results constitute the first terrestrial evidence
of the solar neutrino anomaly. The observation of $\bar{\nu}_e$ disappearance
reinforces the interpretation of the previous neutrino data through $\nu_e
\rightarrow \nu_{\mu,\tau}$ oscillations. The combined analysis of the KamLAND
spectrum with the existent solar and terrestrial data already exclude some
portions of the allowed region in the ($\Delta m_{12}^2, s_{12}^2$)-plane for
the LMA solution
\cite{Barger:2002at,Fogli:2002au,Maltoni:2002aw,Bahcall:2002ij}. One of the
consequences of these global analyses is the splitting of the LMA region in two
sub-regions. In Fig.~\ref{fig4} we plot the allowed regions for the effective
neutrino mass parameter $m_{ee}$ under the initial assumption $\det\,(M_\nu)=0$
and taking as an example the $\pm 1\sigma$ estimates $\Delta m_{12}^2 \simeq
(7.3\pm 0.8)\times 10^{-5}\,\rm{eV}^2$, $s_{12}^2 \simeq 0.315 \pm 0.035$ and
$s_{13} \lesssim 0.13$ given in Ref.~\cite{Fogli:2002au}. In this case there is
no overlap between the two regions corresponding to the hierarchical and
inverse-hierarchical spectra of the light neutrinos. Moreover, it is noticeable
that cancellations on $m_{ee}$ are no longer present. This example also reveals
the importance of future neutrino data in removing the ambiguities of the
neutrino mass matrix.

\section{Conclusions}

In order to remove the ambiguities in the reconstruction of the neutrino mass
matrix from input data, we have proposed in this letter to use the weak-basis
independent condition that the determinant of the effective neutrino mass
matrix vanishes. Since the condition $\det\,(M_\nu)=0$ gives two additional
constraints on the parameters, the resulting mass matrix has only 7 independent
quantities which can be fully determined by future feasible neutrino
experiments.

One may wonder about the stability of our ansatz $\det\,(M_\nu)=0$ under
radiative corrections. In the basis where the neutrino mass matrix is diagonal,
it is easy to show that the vanishing eigenvalue in $M_\nu$ remains zero as
long as supersymmetry is exact (SUSY nonrenormalization theorem). Thus, a
nonvanishing value for $m_1$ is generated only after SUSY-breaking effects are
switched on. The leading contribution comes from two-loop diagrams with
internal $m_2$ or $m_3$ insertions. One can show that this small induced mass
has a negligible effect on our condition $\det\,(M_\nu)=0$ unless $\tan \beta$
is very large.

We have focused our analysis on the correlation between the Majorana
CP-violating phase and the various mixing angles. In particular, we have
discussed how future neutrinoless double beta decay experiments could
invalidate the above weak-basis independent condition. In this framework, we
have also illustrated how one could determine the Majorana phases through
``perfect experiments", i.e. assuming that the values of all seven experimental
input parameters are measured. However, we should remark that in the most
general case, the determination of the Majorana phases through neutrinoless
double beta decay experiments can be a difficult task \cite{deGouvea:2002gf}
due to the uncertainties in the nuclear matrix elements involved in the
extraction of $m_{ee}$ \cite{Rodejohann:2002ng}.

Finally, we have also discussed the impact of the recent KamLAND results on the
effective neutrino mass parameter.

\bigskip \textbf{Acknowledgements}

This work was partially supported by {\em Funda\c{c}{\~a}o para a Ci{\^e}ncia e a
Tecnologia} (FCT, Portugal) through the projects CERN/FIS/43793/2001 and CFIF -
Plurianual (2/91). The work of R.G.F. and F.R.J. was supported by FCT under the
grants SFRH/BPD/1549/2000 and \mbox{PRAXISXXI/BD/18219/98}, respectively.

\end{document}